\documentclass{PoS}

\usepackage{amssymb}
\usepackage{fontenc}
\usepackage{times}
\usepackage{mathptmx}
\usepackage{graphicx}
\usepackage{amsmath}

\title{Search for a continuum limit of the PMS phase}

\ShortTitle{Search for a continuum limit of the PMS phase}

\author{\speaker{Venkitesh Ayyar} $^{ab}$ \thanks{Work done in collaboration with Shailesh Chandrasekharan. The material presented here is based upon work supported by the U.S. Department of Energy, Office of Science, Nuclear Physics program under Award Number DE-FG02-05ER41368. This research was done using resources provided by the Open Science Grid, which is supported by the National Science Foundation and the U.S. Department of Energy's Office of Science.} \\
\llap{$^a$}Department of Physics, Duke University,\\
Durham NC, USA.\\
\llap{$^b$}Department of Physics, University of Colorado,\\
Boulder CO, USA.\\
E-mail: \email{vpa@phy.duke.edu} }

\abstract{Previous studies of a simple four-fermion model with staggered fermions in 3D have shown the existence of an exotic quantum critical point, where one may be able to define a continuum limit of the Paramagnetic Strong Phase (or the PMS phase). We believe the existence of the critical point suggests a new mechanism for generating fermion masses. In this work we begin the search for this quantum critical point in 4D by extending the 3D model to 4D.
Unlike in 3D, now we do find evidence for an intermediate spontaneously broken phase (FM phase) and are able compute the phase boundaries accurately. In terms of the bare coupling,  the width of the intermediate region appears to be quite small.}

\FullConference{34th annual International Symposium on Lattice Field Theory\\
         24-30 July 2016\\
         University of Southampton, UK}

\begin{document}

\section{Introduction}

Lattice field theory models with four-fermion interactions often exhibit interesting phases separated by quantum critical points. Previous studies \cite{bock,anna1,anna2,lee1} of lattice Yukawa and four-fermion models have shown very interesting phase structures. Some of these models have shown the existence of an exotic symmetric phase at strong couplings called the Paramagnetic Strong phase or PMS phase, where fermions are massive without the formation of any bilinear condensates. 

The conventional mechanism for fermion mass generation (which is realized in the strong sector of the Standard Model), involves the Spontaneous Symmetry Breaking (SSB) of chiral symmetries. This is signalled by a non-zero fermion bilinear condensate. In contrast, in lattice models with this exotic PMS phase, lattice fermions are massive without the formation of any fermion bilinear condensates. Hence, the existence of a continuum limit of this PMS phase would imply the existence of a new mechanism for fermion mass generation without any SSB. In our previous work \cite{3d_prd_rc}, using a lattice four-fermion field theory model in 3 Euclidean dimensions, we showed the existence of a continuum limit of this PMS phase. This has been confirmed by other groups \cite{simon,cenke1,cenke2}. In this work, we look for a similar continuum limit of the PMS phase using the same model in 4 Euclidean dimensions. A detailed account of this study can be found in \cite{4d_paper}.

\section{Motivation}
\begin{figure}[]
\centering
\includegraphics[width=0.55\textwidth]{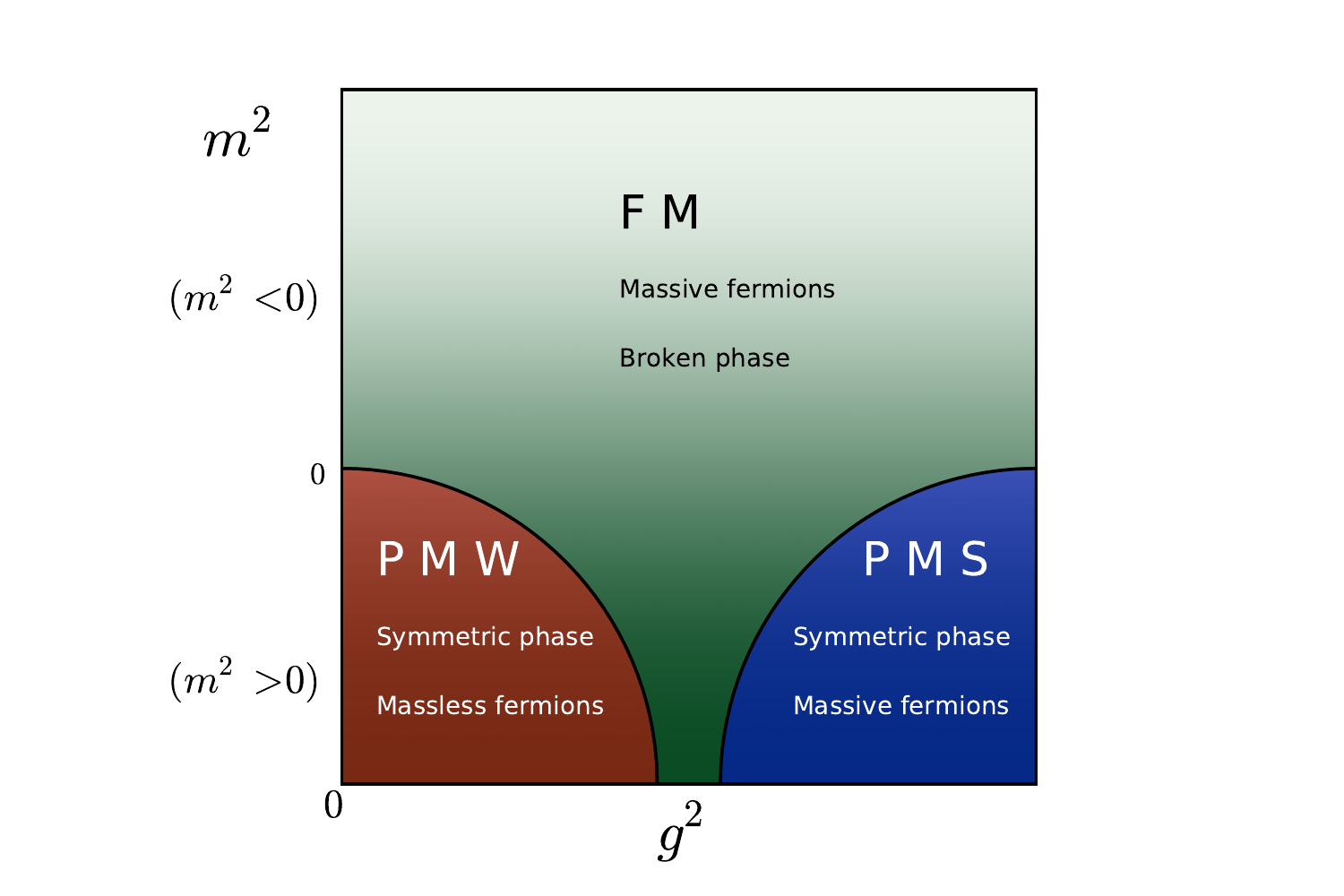}
\caption{The generic phase diagram for some Yukawa models with scalar mass parameter $ m $ and coupling $ g $. It shows the Strong Paramagnetic phase (PMS) phase in addition to the Weak Paramagnetic phase (PMW) and the Ferromagnetic phase (FM).  }
\label{pms_phase} 
\end{figure}

Lattice Yukawa models were studied extensively in the late 1980's and early 1990's. These studies revealed a very interesting phase structure. Most models showed two phases with a symmetric massless fermion phase at weak couplings, called the Weak paramagnetic phase (PMW phase) and a spontaneously broken massive fermion phase at intermediate couplings, called the Ferromagnetic phase (FM phase). However, some models showed another symmetric phase at strong couplings called the Strong paramagnetic phase (PMS phase). This general phase diagram is shown in Fig. \ref{pms_phase}.  

\begin{figure}[tbp]
\centering
\parbox{7cm}{
\includegraphics[width=\linewidth]{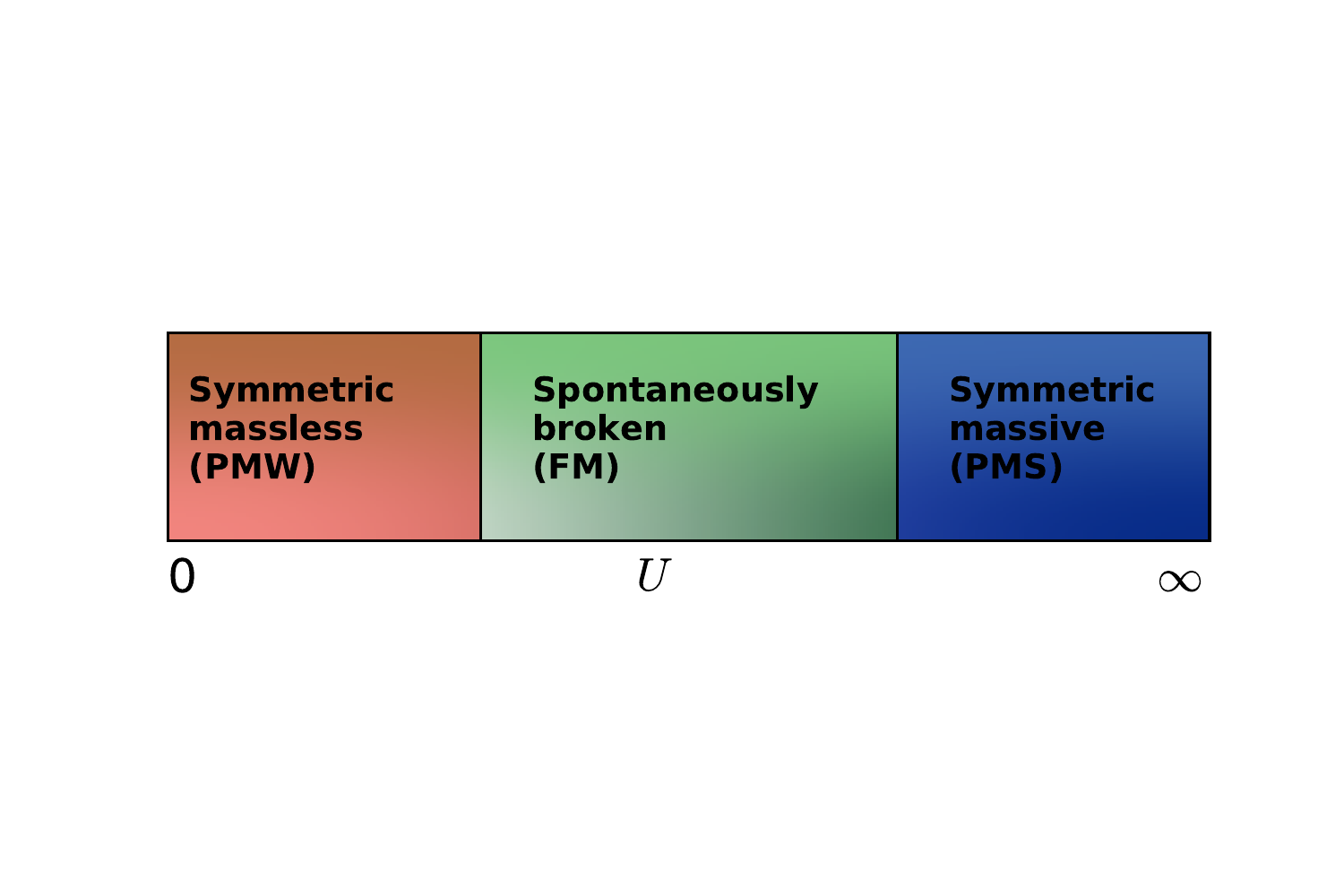}
\caption{\label{4fermion_3phases}The expected phase diagram for four-fermion models with the strong paramagnetic phase (PMS) at strong couplings. }}
\qquad
\begin{minipage}{7cm}
\includegraphics[width=\linewidth]{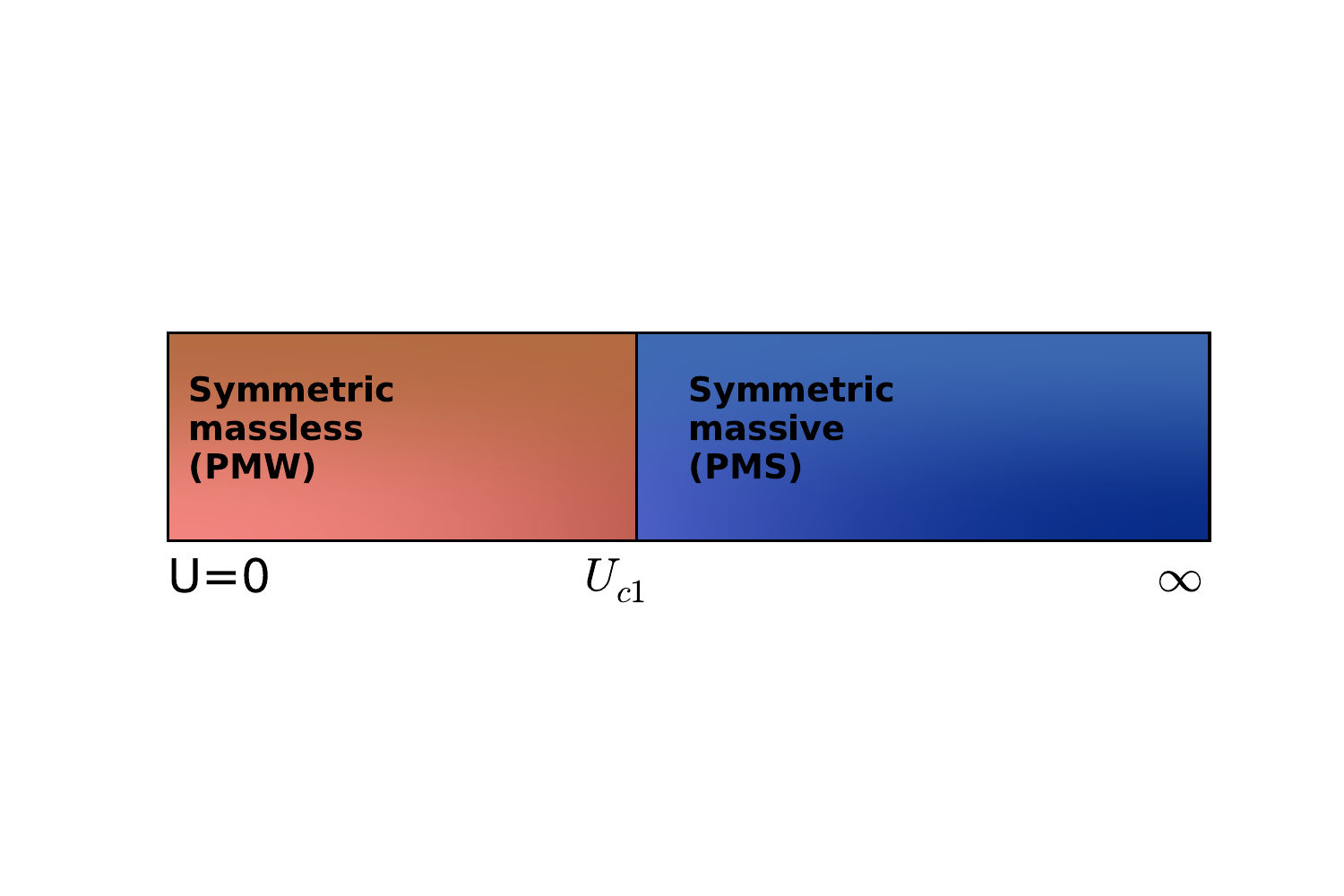}
\caption{\label{4fermion_2phases}The phase diagram for our four-fermion model  with a single second order PMW-PMS phase transition found in 3D.}
\end{minipage}
\end{figure}

What makes the PMS phase interesting is that fermions in this phase acquire a mass without the formation of any fermion bilinear condensates. Hence, lattice models containing a PMS phase are interesting in their own right. Existence of a continuum limit of the PMS phase would be very interesting in particle physics since this would provide a new mechanism for fermion mass generation without Spontaneous Symmetry Breaking. However, since Yukawa models are computationally more expensive due the presence of scalar fields coupling to the fermions, previous studies \cite{lee2} could only be performed on very small lattice sizes. It can be argued that a four-fermion model can capture the same physics as these Yukawa models, along a horizontal line with $ \kappa =0 $. Hence for such a four-fermion model, one expects a phase diagram given in Fig. \ref{4fermion_3phases}. Since four-fermion models are easier to study due to techniques like the Fermion Bag approach \cite{fbag1}, in this work, we study such a four-fermion model. 

If we find the intermediate FM phase to be absent and the PMW-PMS phase transition to be second-order, then this would imply the existence of a continuum limit of the PMS phase. Such a continuum theory would exhibit fermion mass generation without any Spontaneous Symmetry Breaking. In our previous study in 3D, we did find such a 2nd order critical point. This scenario is depicted in Figure \ref{4fermion_2phases}. In this work, we look for such a critical point in 4D.

\section{Our Model}
We study four flavors of massless reduced staggered fermions, interacting via an on-site four-fermion interaction term. This is equivalent to two flavors of reduced staggered fermions as described before \cite{prev_proc}. The action for our model is given by
\begin{equation}
S = S_0 -U \sum_x \left( \psi_{x,1} \psi_{x,2} \psi_{x,3} \psi_{x,4} \right) \label{act},
\end{equation}
where $ S_0 $ is the free reduced staggered action and is given by
\begin{equation}
S_0 = \frac{1}{2} \sum_{i=1}^4 \sum_{x,y} \psi_{x,i} M_{x,y} \psi_{y,i} .
\end{equation}
Here, $ \psi_{x,i}, \ i=1,2,3,4 $ are four independent Grassmann valued fields that represent the four flavors of reduced staggered fermions,  $ x $ represents the sites of a hypercubic lattice, and $ M $ is the well-known free staggered fermion matrix.

In addition to the usual discrete space-time symmetries, the action also has an $ SU(4)$ flavor symmetry. In this work, we explore the order parameter $ \psi_{x,a} \psi_{x,b} $ that breaks the $SU(4)$ symmetry. The observables we measure are:
\begin{eqnarray}
\rho_m= \frac{U}{V} \sum_x {\langle  \psi_{x,1} \psi_{x,2} \psi_{x,3} \psi_{x,4} \rangle} ,\nonumber \\
 \chi_1 = \frac{1}{2} \sum_{x} \ \langle \psi_{0,1} \psi_{0,2} \psi_{x,1} \psi_{x,2} \rangle , \nonumber\\
 \chi_2 = \frac{1}{2} \sum_{x} \ \langle \psi_{0,1} \psi_{0,2} \psi_{x,3} \psi_{x,4} \rangle , \label{obs}
\end{eqnarray}
where $\rho_m $ is a four-point condensate, while $ \chi_1 $ and $ \chi_2 $ are the two susceptibilities corresponding to the order parameter $ \psi_{x,a} \psi_{x,b} $. If the $ SU(4)$ symmetry is broken spontaneously, then a condensate $ \Phi $ forms, and hence the susceptibilities will scale with the volume i.e $ \chi_{1,2} \sim \Phi^2 L^4 $.

\section{Computational Approach}
The conventional approach to solve such four-fermion models is by introducing a scalar auxiliary field and converting it to a bosonic problem. In this work, we use an alternate approach called the Fermion Bag approach \cite{fbag1}. This method deals directly with the fermionic degrees of freedom. It can be applied to a certain class of fermionic models and it has been found to be very efficient in solving these models. Application of the Fermion Bag approach to our model has been described in detail in \cite{4d_paper}.

Our aim is to explore the phase diagram of this model. At weak couplings, since the four-fermion coupling $ U $ is irrelevant, we expect massless fermions. At strong couplings, it can be argued using the fermion bag approach \cite{3d_prd} that all two point correlations must decay exponentially. Hence, we expect massive fermions without any fermion bilinear condensates (PMS phase). We would like to know whether there exists an intermediate spontaneously broken phase (FM phase) at intermediate couplings.

\section{Results}
\begin{figure}[tbp]
\centering
\parbox{7cm}{
\includegraphics[width=\linewidth]{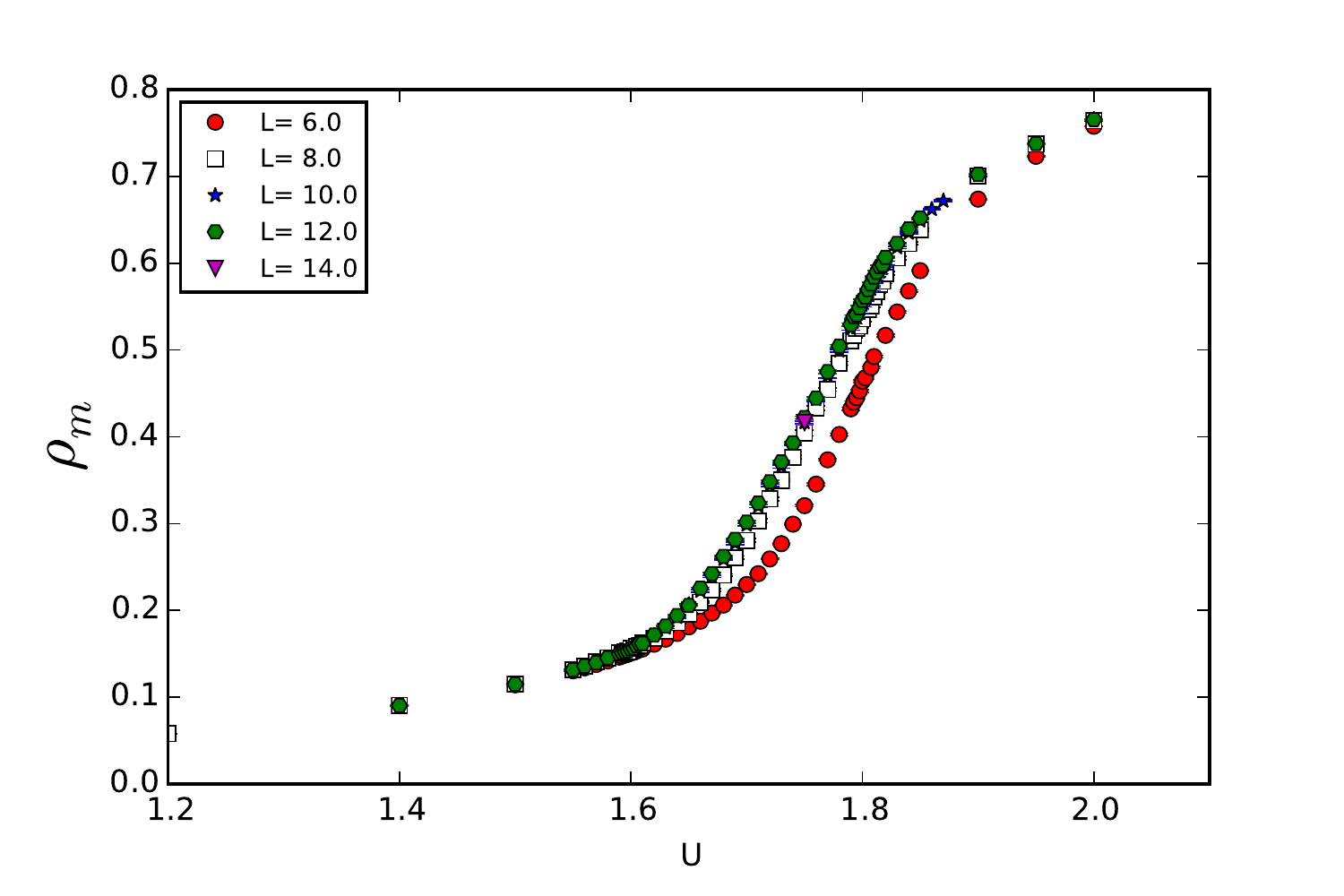}
\caption{\label{rho} Variation of the four-point condensate $ \rho_m $ with coupling $ U $. $ \rho_m$ increases with coupling $ U $, rising sharply near $ U = 1.75 $, without showing any discontinuities. }}
\qquad
\begin{minipage}{7cm}
\includegraphics[width=\linewidth]{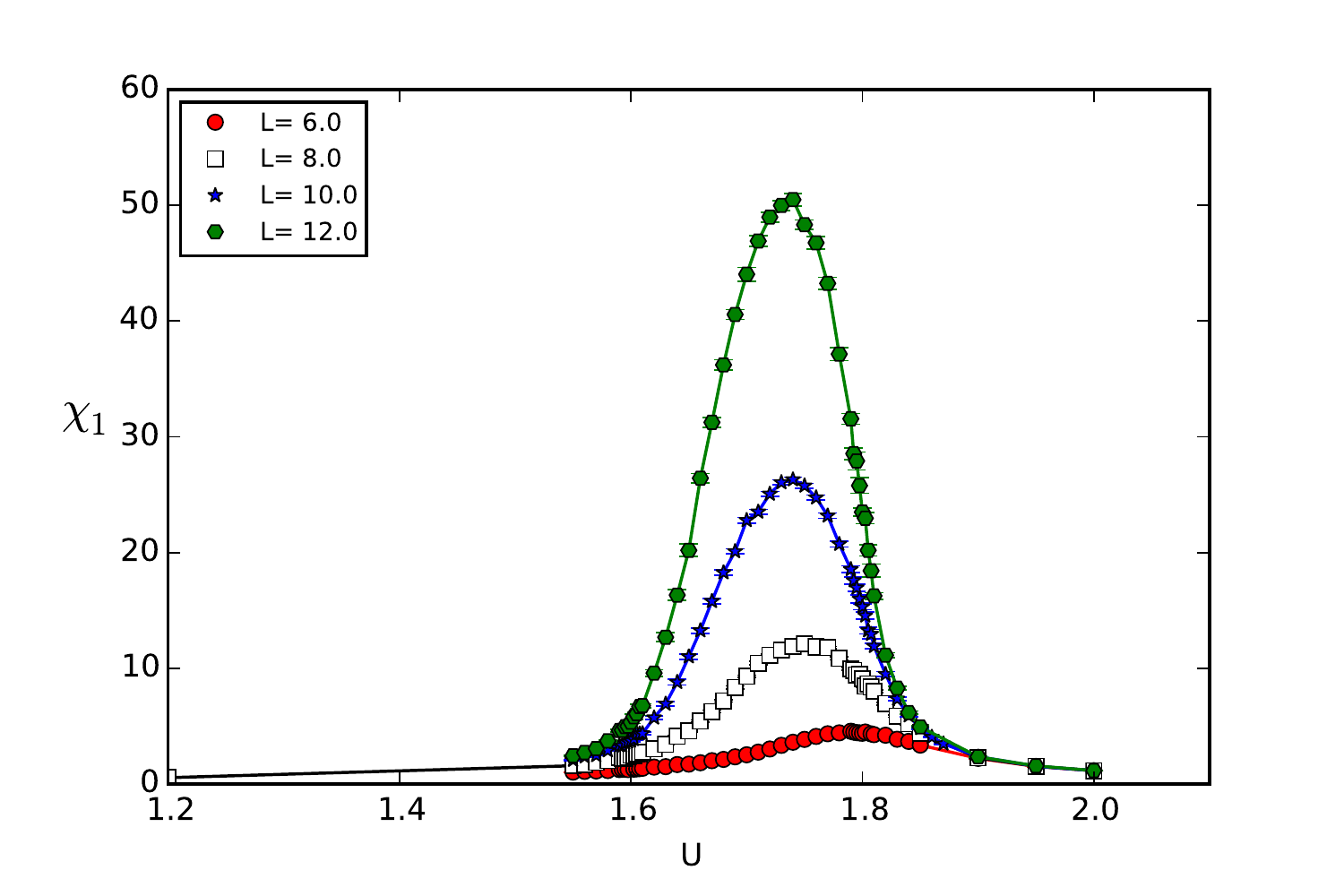}
\caption{\label{chi1_vs_u} Variation of the susceptibility $ \chi_1 $ with coupling $ U $ for various lattice sizes. $\chi_1$  increases with $ U $ to reach a maximum for intermediate $ U $ and then decreases.}
\end{minipage}
\end{figure}

We performed calculations on symmetrical lattices upto size $ 12^4$ with anti-periodic boundary conditions in all directions. The behavior of the observable $ \rho_m $ as a function of the coupling $ U $ for various lattice sizes $ L $  is shown in Fig. \ref{rho}. Being a four-point condensate, any discontinuity in $ \rho_m $ can be a sign of a first-order phase transition. The smooth nature of $ \rho_m $ seen in Fig. \ref{rho} seems to hint at the absence of any first-order transitions.

\begin{figure}[tbp]
\centering
\parbox{7cm}{
\includegraphics[width=\linewidth]{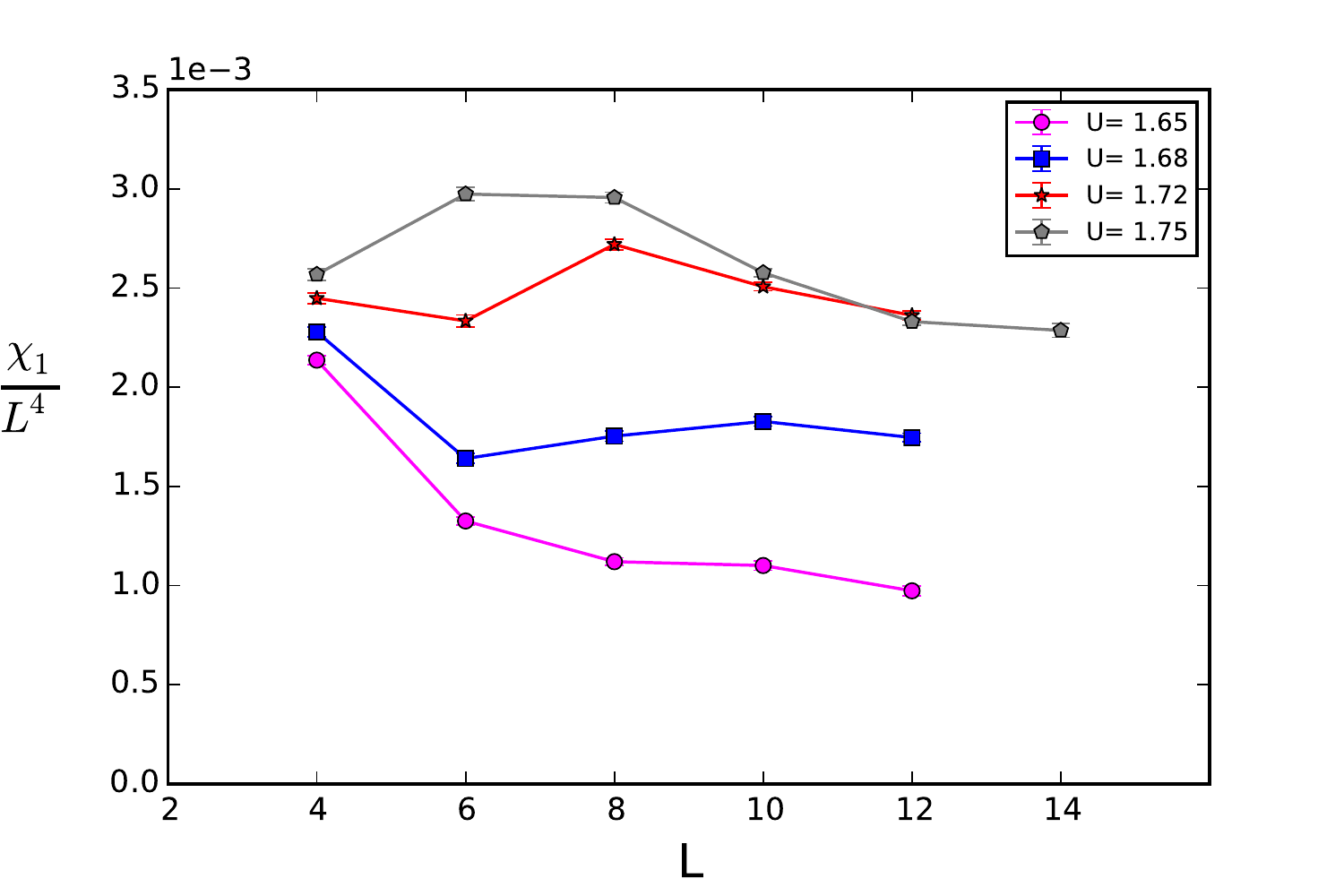}
\caption{\label{chi1_vs_L} The saturation of $ \chi_1 / L^4 $ as lattice size $ L $ increases for various intermediate couplings. This implies the formation of a fermion bilinear condensate at intermediate $ U $.}}
\qquad
\begin{minipage}{7cm}
\includegraphics[width=\linewidth]{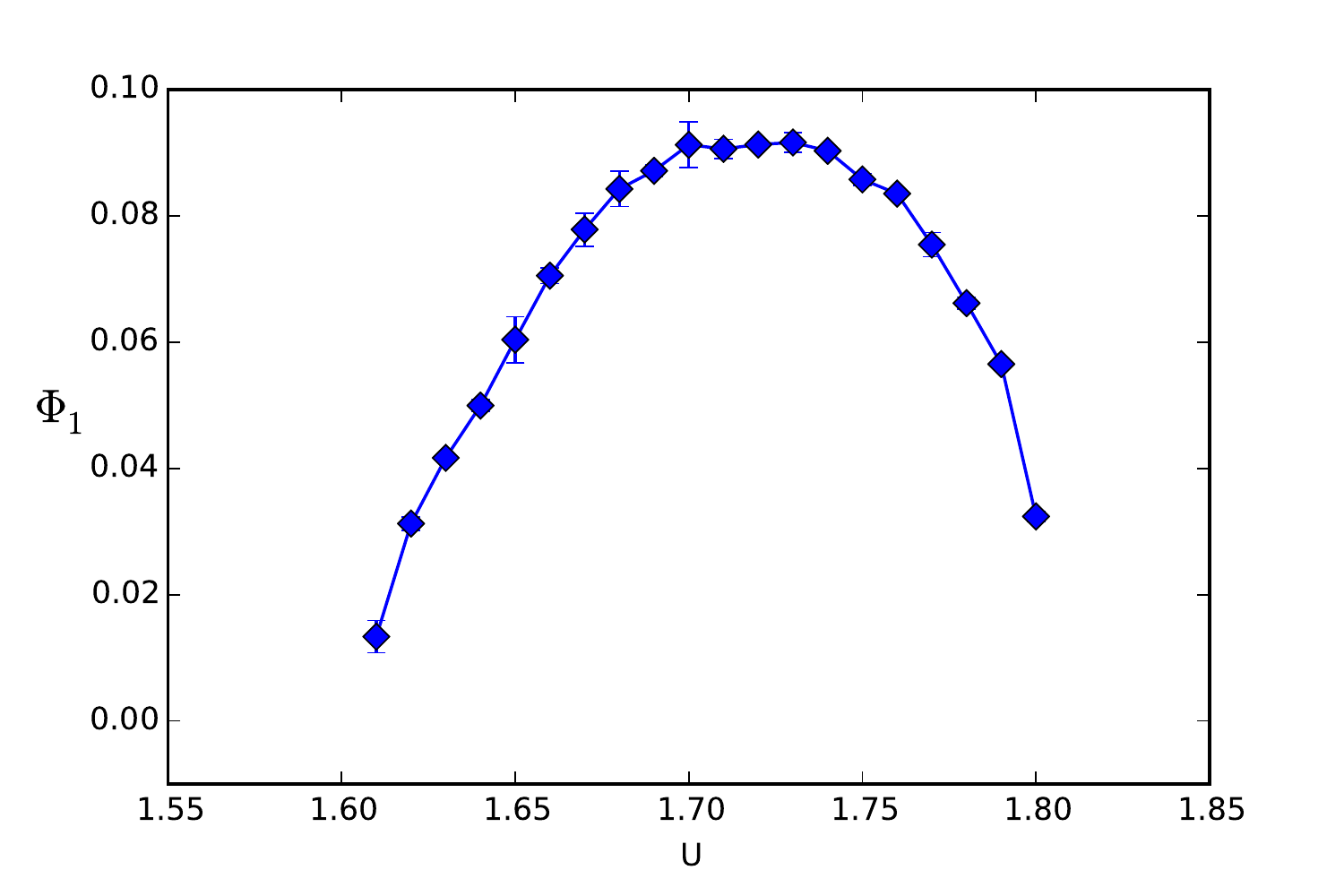}
\caption{\label{chi1_condensate} Plot of the condensate forming in the intermediate region from $ U=1.60$ to $ 1.80$. This is the FM phase.}
\end{minipage}
\end{figure}

To explore the phase diagram further, we look at the behavior of the susceptibilities. Fig. \ref{chi1_vs_u} shows the behavior of $ \chi_1 $ as a function of the coupling $ U $. It can be seen that the susceptibility $ \chi_1 $ increases with $ U $, reaches a maximum at intermediate $ U $ and then decreases at large $ U $. It is clear that at intermediate couplings, $ \chi_1 $ increases sharply with lattice size $L$.

As discussed before, for the existence of fermion bilinear condensates, the susceptibility must grow with the volume $ L^4$. To check the presence of a condensate in the intermediate region, we look at the variation of $ \chi_1 / L^4 $ as a function of $ L $ for some intermediate couplings in Fig. \ref{chi1_vs_L}. It can be seen that the curve saturates as $ L $ increases. This implies that the susceptibility $ \chi_1 $ grows as $ L^4 $ in the intermediate region and hence fermion bilinear condensates form there.
To extract the condensate $\Phi$, we performed a fit of $\chi_1 $ to the form $ \chi_1 = \frac{1}{4} \Phi^2 L^4 + \frac{1}{2} b_1 L^2 $. The variation of this condensate $\Phi $ with coupling $ U $ is shown in Fig. \ref{chi1_condensate}. It is clear that the condensate forms in the intermediate region between $ U=1.60$ and $ U=1.80$. A similar analysis with the susceptibility $ \chi_2 $ gives the same results.

\begin{figure}[tbp]
\centering
\includegraphics[width=0.55\textwidth]{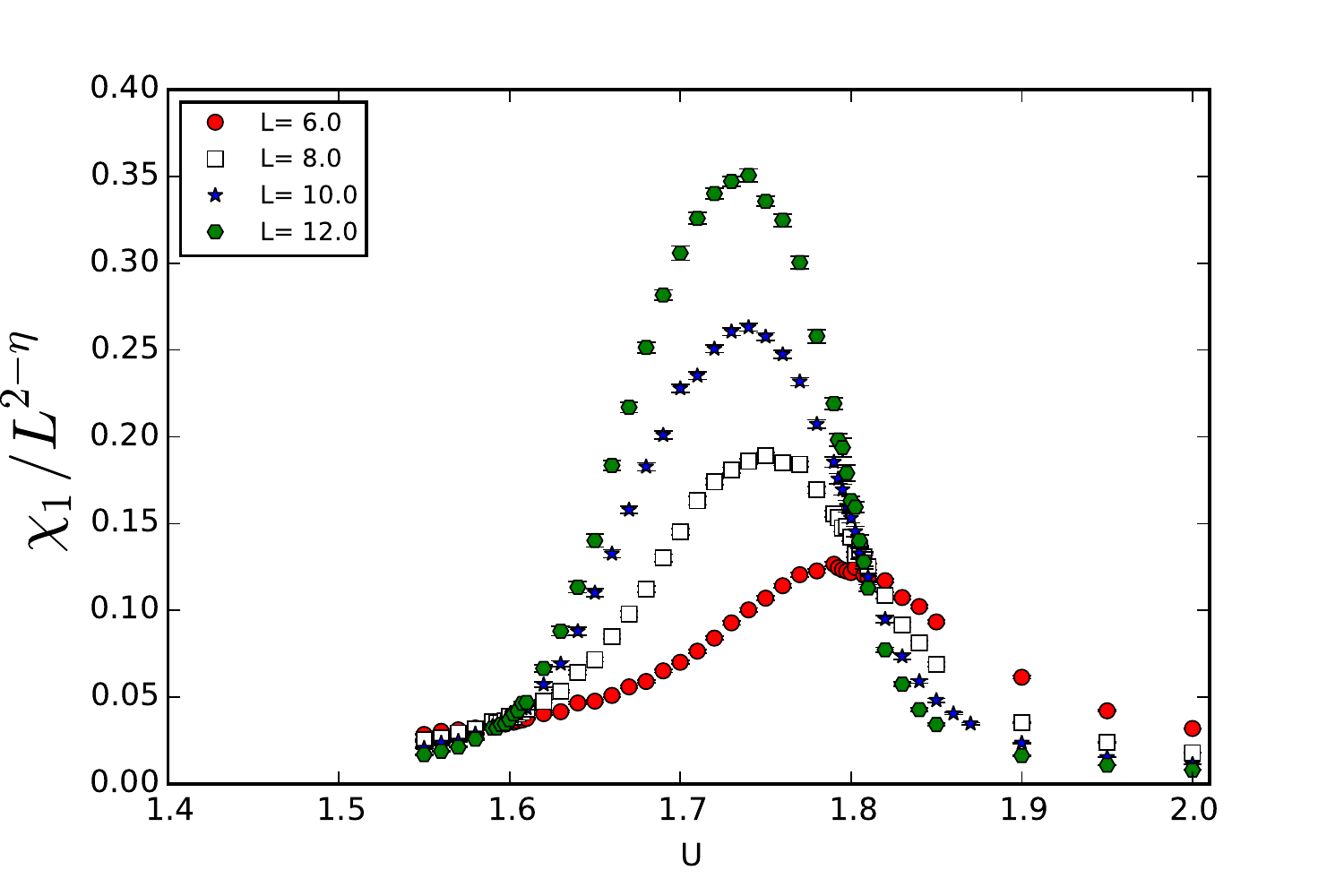}
\caption{\label{crit_scaling} The curves $ \chi_1 / L^{2-\eta}$ vs $ U $ with the mean field value of $ \eta = 0 $ for various lattice sizes $ L $ intersect at two points, $ U_{c1} = 1.60 $ and $ U_{c2} = 1.80 $. This implies that the PMW-FM and FM-PMS transitions are second order with mean field exponents.}
\end{figure}

The presence of an intermediate FM phase implies the existence of two phase transitions: a PMW-FM and an FM-PMS transition. We would like to understand the nature of these phase transitions. From the theory of second-order phase transitions, we know that $ \chi \sim L^{2-\eta}$ near a second order critical point. Making an ansatz that these two transitions are second-order with mean-field exponents $ \eta = 0 $ and $ \nu = 0.5 $, we plot $ \chi_1 / L^{2-\eta} $ as a function of coupling $U$ in Fig. \ref{crit_scaling}. The curves intersect at the two points $ U_{c1} = 1.60 $ and $ U_{c2} = 1.80 $. This confirms our suspicion that the two transitions are both second-order with the mean field exponents.

\section{Conclusions}
Thus, our study of a four-fermion lattice model with 4 flavors of reduced staggered fermions in 4 Euclidean dimensions shows the presence of a 3 phase structure, with a Ferromagnetic (FM) phase at intermediate couplings sandwiched by a weak Paramagnetic (PMW) phase at weak couplings, a strong Paramagnetic (PMS) phase at strong couplings. In the PMS phase, the fermions are massive without the formation of any fermion bilinear condensates. Our analysis shows that the two transitions are both second-order with mean field exponents $\eta=0$ and $ \nu = 0.5$.

Even though the presence of the intermediate FM phase makes this model less interesting, the presence of the PMS phase is interesting in its own right. The actual mode for fermions to acquire a mass in this PMS phase is not clear. One possibility is that the fermions may be acquiring a mass via the formation of four-fermion condensates. At strong couplings, non-perturbative dynamics could result in three fundamental fermions forming a bound state to give a composite fermion. This composite fermion could couple to a fundamental fermion, and such a mass term would resemble a four-fermion condensate in microscopic theory. This possibility has been proposed before \cite{eichton,golterman}.

While the presence of the intermediate FM phase precludes the existence of a PMW-PMS second order phase transition in this model in 4D, the width of this FM phase is found to be quite small. This raises an interesting possibility: it might be possible to eliminate this phase in an enhanced coupling space by tuning the couplings. We intend to explore this possibility in future work by adding higher order terms.


\begin{thebibliography}{99}

\bibitem{bock}
W. Bock, A. K. De, K. Jansen, J. Jersak, T. Neuhaus, and J. Smit, \\
\emph{Phase diagram of a lattice SU(2) $ \times $ SU(2) scalar-fermion model with naive and Wilson fermions},\\
Nucl.Phys. B344, 207 (1990).

\bibitem{anna1}
A. Hasenfratz, T. Neuhaus,\\
\emph{Non-perturbative study of the strongly coupled scalar-fermion model},\\
Phys.Lett. B, Vol 220, Issue 3.(1989)

\bibitem{anna2} 
A. Hasenfratz, W.-q. Liu, and T. Neuhaus,\\
\emph{Phase Structure and Critical points in a scalar-fermion Model},\\
Phys.Lett. B236, 339 (1990).

\bibitem{lee1}
I.-H. Lee, J. Shigemitsu, and R. E. Shrock, \\
\emph{Lattice study of a Yukawa theory with a real scalar field}, \\
Nucl.Phys. B330, 225 (1990).

\bibitem{lee2}
I.-H. Lee, J. Shigemitsu, and R. E. Shrock, \\
\emph{Study of different lattice formulations of a Yukawa model with a real scalar field }, \\
Nucl.Phys. B334, 265 (1990).

\bibitem{3d_prd_rc}
V. Ayyar and S. Chandrasekharan, \\
\emph{Origin of fermion masses without spontaneous symmetry breaking}, \\
Phys Rev D 93, 081701(R), (2016).


\bibitem{simon}
Simon Catterall,\\
\emph{Fermion mass without symmetry breaking},\\
JHEP 1601 (2016) 121.

\bibitem{cenke1}
K. Slagle, Y. You and Cenke Xu, \\
\emph{Exotic quantum phase transitions of strongly interacting topological insulators}, \\
Phys. Rev. B 91, 115121 (2015).

\bibitem{cenke2}
Yuan-Yao He, Han-Qing Wu, Yi-Zhuang You, Cenke Xu, Zi Yang Meng, Zhong-Yi Lu, \\
\emph{Quantum critical point of Dirac fermion mass generation without spontaneous symmetry breaking},\\
arXiv:1603.08376.

\bibitem{4d_paper}
V. Ayyar, S. Chandrasekharan, \\
\emph{Fermion masses through four-fermion condensates}, \\
JHEP, doi:10.1007/JHEP10(2016)058, (2016). 

\bibitem{fbag1} 
S. Chandrasekharan, \\
\emph{Fermion bag approach to lattice field theories },\\
Phys. Rev. D 82, 025007 (2010).

\bibitem{prev_proc}
V. Ayyar, \\
\emph{Exotic Quantum Critical Points with Staggered Fermions}, \\
Proceedings of Science, Lattice 2015.

\bibitem{3d_prd}
V. Ayyar and S. Chandrasekharan, \\
\emph{Massive fermions without fermion bilinear condensates}, \\
Phys Rev D 91, 065035, (2015).

\bibitem{eichton}
E. Eichten and J. Preskill,\\
\emph{Chiral gauge theories on the lattice},\\
Nucl. Phys. B268, 179 (1986).

\bibitem{golterman}
M. F. L. Golterman, D. N. Petcher, and E. Rivas,\\
\emph{Absence of chiral fermions in the Eichten-Preskill model},\\
Nucl. Phys. B395, 596 (1993)

\end{thebibliography}
\end{document}